# Prediction of inorganic superconductors with quasi-one-dimensional crystal structure

**L M Volkova and D V Marinin**

Institute of Chemistry, Far Eastern Branch of the Russian Academy of Sciences
690022 Vladivostok, Russia

E-mail: volkova@ich.dvo.ru

**Abstract**
Models of superconductors having a quasi-one-dimensional crystal structure based on the convoluted into a tube Ginzburg sandwich, which comprises a layered dielectric–metal–dielectric structure, have been suggested. The critical crystal chemistry parameters of the Ginzburg sandwich determining the possibility of the emergence of superconductivity and the $T_c$ value in layered high-$T_c$ cuprates, which could have the same functions in quasi-one-dimensional fragments (sandwich-type tubes), have been examined. The crystal structures of known low-temperature superconductors, in which one can mark out similar quasi-one-dimensional fragments, have been analyzed. Five compounds with quasi-one-dimensional structures, which can be considered as potential parents of new superconductor families, possibly with high transition temperatures, have been suggested. The methods of doping and modification of these compounds are provided.

## 1. Introduction

For the first time, the notion on the possibility of the emergence of high-temperature superconductivity (HTSC) in quasi-one-dimensional systems was put forward by Little in 1964 [1]. His model system comprises a molecule consisting of two parts: first, a long chain called the 'spine', in which electrons fill various states and may or may not form a conducting system, and, second, a series of arms or side chains attached to the spine. However, up to recently, the model has not yet been realized. The same year Ginzburg suggested a quasi-two-dimensional system – metal film with dielectric layers from both sides (sandwich – layered structure: dielectric – metal – dielectric (*I/S/I*)) – as high-temperature superconductor [2]. Thereafter, Katz demonstrated [3] that layered chemical compounds with dramatically different conductivities of adjacent layers could be considered as stacks made of the above sandwiches. Although one have not yet managed to create such superconducting systems artificially, in these very two-dimensional systems (complex copper oxides) high-temperature superconductivity was discovered for the first time by Bednorz and Müller [4]. According to Ginzburg [5], these two model systems are based on the common notion consisting in a spatial separation of the conductivity (superconductivity) region and the dielectric region responsible for the electron polarization.

However, the earlier performed studies [6-9] of correlations between $T_c$ and crystal chemistry parameters ratios in superconducting fragments (sandwiches of different families of superconducting cuprates) demonstrated that the maximum value of $T_c$, which could be attained on the basis of the $CuO_2$ plane, would not exceed ~160 K independently of the number of such planes in a superconductor. So we came to the idea of creating models of superconductors with a quasi-one-dimensional crystal based on the Ginzburg sandwich that could have higher $T_c$.

In the present work, we will outline crystal chemistry parameters of the Ginzburg sandwich determining the possibility of the emergence of superconductivity and $T_c$ in layered high-$T_c$ cuprates, which could have the same functions in quasi-one-dimensional fragments – tubes of the sandwich type. Taking into account all these parameters, we will suggest models of superconductors with a quasi-one-dimensional structure based on the Ginzburg sandwich convoluted into a tube. We will demonstrate that in crystal structures of some well-known low-temperature superconductors one can mark out similar quasi-one-dimensional fragments of the sandwich type. Then we will identify them in the Inorganic Crystal Structure Database (ICSD) (version 1.8.2, FIZ Karlsruhe, Germany, 2012-1) and present 5 compounds with a quasi-one-dimensional structure – potential parents of new families of superconductors having, possibly, high transition temperatures.

## 2. Crystal chemistry prediction criteria

To predict new superconductors, it is important to know the role of crystal chemistry factors in creating conditions for the emergence of high temperatures of transition into the superconducting state ($T_c$). The earlier performed studies [6-9] of high-temperature superconducting cuprates attaining, for the moment, the highest transition temperatures $T_c$, enabled us to reveal these factors. Below we will explain it in more detail. In cuprates with several $CuO_2$ layers the perovskite-like layer $A_{n+1}(CuO_2)_n$ is divided, in accordance with the number of $CuO_2$ planes ($n$), to $n$ elementary fragments – $A_2CuO_2$ sandwiches (figure 1). Even at equal d(Cu-Cu) distances they could differ in the concentration of charge carriers (number of holes (p) in $CuO_2$ layers as calculated per one copper atom), the distances between the $CuO_2$ plane and those of A-cations, and element compositions of A-cations planes. The main difference of the structure of cuprate HTSC from that of perovskite type of other elements consists in the fact that $CuO_2$ planes emerge between positively charged planes of A-cations as a result of the moving off (or absence) of apical oxygen atoms from the $CuO_2$ plane by larger distances than A-cations planes due to a strong Jahn-Teller effect.

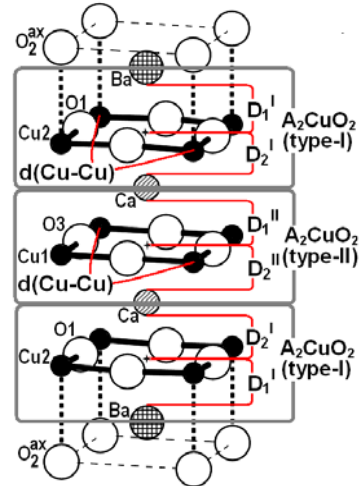

**Figure 1.** Perovskite-type layer $A_{n+1}(CuO_2)_n$ (n = 3) in the crystal structure of $HgBa_2Ca_2Cu_3O_8$. The superconducting fragments (sandwiches) are highlighted.

We revealed [6-9] an empirical correlation $T_c(J)$ between $T_c$ and the value of the ratio ($J$) of the distance $d$(Cu-Cu) between copper atoms along the diagonal direction of the $CuO_2$ plane to the sum of effective distances ($D_1+D_2$) from the $CuO_2$ plane to the surface of two adjacent layers of A cations in superconducting fragment (sandwich) $A_2CuO_2$:

$$J = d(Cu\text{-}Cu)/(D_1+D_2) \qquad (1)$$

The effective distances $D_1$ and $D_2$ were calculated with taking into account the effect of the field and outer sandwich planes on the carriers:

$$D = S[d(CuO_2 - A) - R_A(Z_A/2)] \qquad (2)$$

where $d(CuO_2\text{-}A)$ is the distance from $CuO_2$-plane to plane of A-cations; $R_A$ is the radius of A-cation, which content is maximal, $Z_A/2$ – the dimensionless coefficient to take into account the electric field of the A-cation charge (it is the ratio of charge A-cation to that of Ca cation); $S$ is the deviation coefficient of parameters of doping cations from those of A-cation that forms the plane:

$$S \geq 1, \quad S = \overline{R(Z/2)}/R_A(Z_A/2) \text{ or}$$
$$S = R_A(Z_A/2)/\overline{R(Z/2)}$$

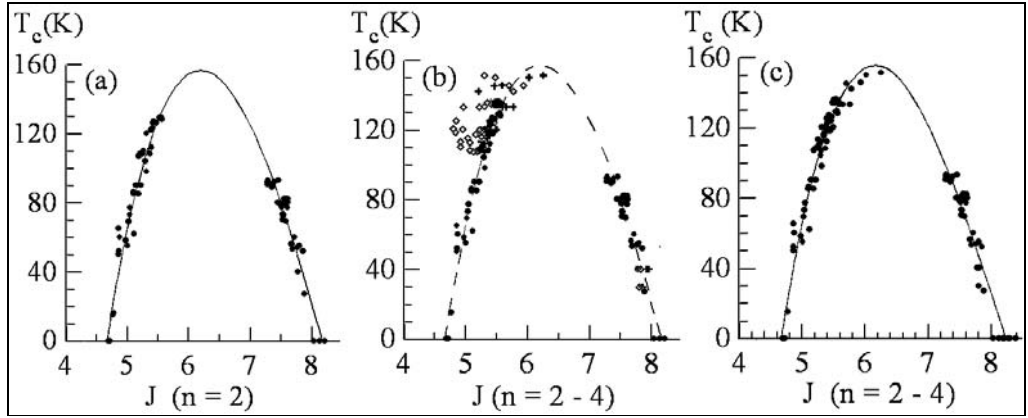

**Figure 2.** $T_c$ as a function of $J$ ($J = d$(Cu-Cu)/($D_1+D_2$)) in high-$T_c$ cuprates: (a) with only one type (external) of fragments $A_2CuO_2$, (b) cuprates with two types of fragments (external (+) and internal (◊)) were added to cuprates with one type of fragments, (c) fragments with higher $T_c$ were selected from cuprates with two types of fragments. Reproduced with permission from [6]. Copyright 2007 Nova Science Publishers, Inc.

Here $\overline{R(Z/2)}$ is a generalized value, characterizes the plane of A-cations:

$$\overline{R(Z/2)} = m_1 R_{A_1}(Z_{A_1}/2) + ... m_n R_{A_n}(Z_{A_n}/2)$$

where $m_n$ is the content of $A_n$-cation in plane, $R_{A_n}$ is the radius, $Z_{A_n}/2$ – is the dimensionless coefficient to take into account the electric field of the $A_n$-cation charge. One should mention that the A cations radii were taken from [10] with the coordination number 8.

The correlation $T_c(J)$ is of a general character for different families of superconducting cuprates with several $CuO_2$ planes (n = 2 – 4) and any cation composition and doping degree. The $T_c(J)$ curve (figure 2(c)) was built without using any normalizing coefficients for 132 compounds (among them, with the number of $CuO_2$ planes: n = 2 – 94, n>2 – 38). In graphic terms, the $T_c(J)$ correlation is close to parabola; however, the best approximation (97%) of this correlation is provided by a third-degree polynomial equation.

In the $T_c(J)$ correlation, the number of holes is taken into account indirectly through the value of the distance $d$(Cu-Cu), to which it is virtually inverse proportional, since the oxygen atoms are located immediately in the copper atoms plane or slightly deviate from it. According to the estimation of the value of charges of copper (q) on the bond valence sum (BVS) of copper (BVS = $\Sigma_i \exp[(1{,}679-R_i)/0{,}37]$ [11], where $R_i$ – the Cu-O distance), the copper atoms charge and, therefore, the number of holes (p = q-2) in external $CuO_2$ planes is higher (by the fraction contributed from the apical oxygen atom) than that in internal planes. However, our attempts to build a universal (general) correlation for all HTSC families with taking into account the increase of the number of holes in external $CuO_2$ planes was not successful. So one can conclude that $T_c$ depends on the number of holes emerging due to contributions from oxygen atoms located exclusively in the $CuO_2$ plane.

According to the $T_c(J)$ correlation, the transition temperature $T_c$ does not depend on the number of $CuO_2$ planes in a superconductor that contradicts the generally accepted opinion. At n>2 the crystal chemistry parameters of sandwiches formed by external (type I) and internal (type II) $CuO_2$ planes are not equivalent that yields different values of $J$ and, therefore, $T_c$ (figure 2(b)). The transition temperature $T_c$ of the compound having several sandwiches is the individual $T_c$ of one of them, namely, the sandwich, whose parameters ($d$(Cu-Cu) distances and sum of effective $D_1$ and $D_2$ distances) are close to optimal ones and $T_c$ is higher. Figure 2(a) shows the dependence of $T_c(J)$ built for cuprates (n = 2) having identical sandwiches of type I, in figure 2(b) they are complemented with cuprates (n > 2) with different types (I and II) of sandwiches, whereas in figure 2(c), in the presence of two sandwich types, those having the $J$ ratio corresponding to higher $T_c$ are selected.

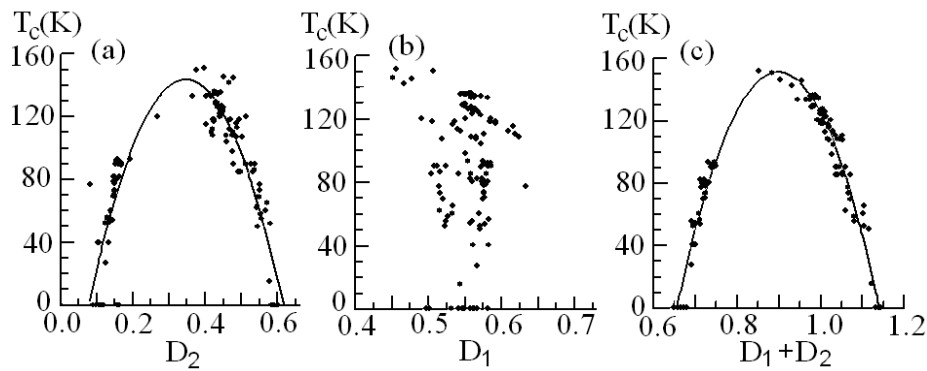

**Figure 3.** $T_c$ as a function of (a) the effective distance $D_2$ between the $CuO_2$ plane and the layer of A cations that does not contain oxygen atoms in superconducting cuprates with the number of layers n>1, (b) effective distance $D_1$ from the $CuO_2$ plane to the external plane of A cations and (c) the sum $D_1+D_2$ in high-$T_c$ HTSC cuprates formed by some $CuO_2$ planes (n = 2-4). Reproduced with permission from [6]. Copyright 2007 Nova Science Publishers, Inc.

One should emphasize that in most (28 from 38) of the examined compounds with several types of sandwiches $A_2CuO_2$ the compound $T_c$ is determined by the sandwich formed through the external $CuO_2$ plane. For example, in the superconductor $HgBa_2Ca_2Cu_3O_{8+d}$ [12] at low pressure down to 0,59 GPa ($T_{c,\,exp}$ = 135-136K), the compound $T_c$, according to the curve $T_c(J)$, is that of sandwiches of type I ($T_{c,\,calculat.}^{I}$ = 128-133K, $J^I$ = 5,49 – 5,56), whereas the transition temperature of the sandwich of type II is noticeably lower ($T_{c,\,calculat.}^{II}$ = 120-125 K, $J^{II}$ = 5,40 –5,46). Along with the pressure increase up to 4 GPa ($T_{c,\,exp}$ = 142 K), the compound $T_c$ is determined by the sandwich of type II ($T_{c,\,calculat.}^{II}$ = 147 K, $J^{II}$ = 5,78), whereas the transition temperature of fragments of type I is significantly lower ($T_{c,\,calculat.}^{I}$ = 96 K, $J^I$ = 5,20). Further pressure increase up to 9,2 GPa ($T_{c,\,exp}$ = 151.1 K) returns the leading position to the sandwiches of type I ($T_{c,\,calculat.}^{I}$ = 154 K, $J^I$ = 6,25) and decreases dramatically the transition temperature of the sandwich of type II ($T_{c,\,calculat.}^{II}$ = 107 K, $J^{II}$ = 5,30).

The role of parameters $D_1$ and $D_2$ characterizing the distances between the internal $CuO_2$ plane and the surface of layers of A ions, the degree of heterogeneity in A ions plane surfaces, and the field created by charges of A and doping ions in the sandwich $A_2CuO_2$ in superconductivity emergence and $T_c$ changes is very significant. There exists a nearly parabolic dependence (polynomial approximation 81%, $T_c(D_2)$) (figure 3(a)) of $T_c$ on the effective distance $D_2$ between the $CuO_2$ plane and the layer of A cations that does not contain oxygen atoms in superconducting cuprates with the number of layers n>1. At the same time, the external (type I) sandwich $A_2CuO_2$ does not have a correlation between $T_c$ and the effective distance $D_1$ from the $CuO_2$ plane to the external plane of A cations (figure 3(b)), which could be related to the decrease of the effective positive charge of the external plane of cations due to its proximity to $O_{ax}$ atoms and, as a result, the decrease of the focusing effect on carriers. The correlation between $T_c$ and the sum of $D_1$ and $D_2$ is even closer to parabolic (approximation $T_c(D_1+D_2)$ to polynomial 95%) (figure 3(c)), if one takes 124 sandwiches of type I and 8 sandwiches of type II manifesting the highest $T_c$ in accordance with the $T_c(J)$ correlation in the compounds under study.

According to the $T_c(J)$ correlation, the maximal $T_c$, which can be attained in high-temperature superconducting cuprates, is as low as 155 K ($J$ = 6,15), if one uses the structural data of all 132 compounds under examination. The transition temperature of 162 K ($J$ = 6.15, $T_c$ = 12.1113$J^3$ - 286.406$J^2$ + 2149.29$J$ - 5041), close to the highest one of 164 K [13] attained in Hg 1223 (phase under pressure), is obtained upon using the structural data of 50 compounds investigated with the highest accuracy by means of X-ray and neutron single crystal diffraction methods. As follows from the revealed $T_c(J)$ correlation, it is not possible to significantly increase $T_c$ in superconductors based on the $CuO_2$ copper-oxygen plane. However, it is possible to attain this temperature in cuprates without pressure application, if one creates a sandwich with the ratio value $J = 6,15$. In this case, according to $T_c(J)$ correlations (figure 2(c)), $T_c(D_2)$ and $T_c(D_1+D_2)$ (figure 3(a) and (c), the structural parameters of the sandwich $Ca_2CuO_2$ must be as follows: the distances between copper atoms along the diagonal direction of the $CuO_2$ plane $d$(Cu-Cu) = 5.326 Å; the sum of effective distances from the $CuO_2$ plane to the surfaces of two adjacent layers of Ca cations $D_1 + D_2 = 0.866$ Å, where $D_1 = D_2$; the parameter of the copper square lattice $a = 3.766$ Å; the distance $d$(Cu-O) = 1.883 Å; the distance between the $CuO_2$ and Ca planes 1.553 Å; the number of holes p = 0.305. For the sandwich $Mg_2CuO_2$, the parameters are the same, except for the distance between the $CuO_2$ and Mg planes that must be decreased down to 1.323 Å.

We established similar correlations between $T_c$ and the crystal chemistry parameters ratio in other layered superconductors (diborides $AB_2$ and nickel borocarbides $RNi_2B_2C$) as well [14, 15]. Our preliminary studies of $RFeAsO_{1-x}F_x$, $RFeAsO_{1-x}$ (R = lanthanides) and LaOFeP demonstrate that such a correlation exists in Fe-based superconductors.

Thus, we demonstrated that the critical crystal chemistry parameters of the 'Ginzburg sandwich' determining the possibility of emergence of superconductivity and the $T_c$ value in these materials included: concentration of charge carriers, distances between atoms-charge carriers in the metal layer, parameters of the cross-section size of the space between the metal layer and dielectric layers from both sides, and charge and degree of heterogeneity of dielectric layers. The presence of defects in metal layers, overlapping of the transfer space by covalent bond electrons, and different signs of carriers and dielectric layers charges could yield the decrease of $T_c$ until the superconductivity suppression. One of the reasons of high $T_c$ in superconducting cuprates consists in moving off of binding electrons between copper ions and axial oxygen ones from the $CuO_2$ plane due to the Jahn-Teller effect. The above criteria can be applied to develop models and to search for superconductors with quasi-one-dimensional structure, whose temperature of superconducting transition could be higher than in two-dimensional cuprates.

## 3. Crystal chemistry prediction of superconductors with quasi-one-dimensional crystal structure

Since the Little assumption on the possibility of the existence of HTSC in quasi-one-dimensional systems comprising long conducting molecules with side 'branches' has not yet been corroborated, we have decided on using, for the sake of prediction, a quasi-two-dimensional Ginzburg system with taking into account the found crystal chemistry parameters controlling $T_c$ of high-temperature superconducting cuprates. It is possible to transform the Ginzburg sandwich, consisting of dielectric – metal – dielectric layers, into a quasi-one-dimensional system by convoluting it into a tube. In this way, one would avoid the formation of alternative bonds …M-M…M-M… in the Little model 'spine' that hampers the emergence of superconductivity with high $T_c$.

We believe that on the above basis one can build quasi-one-dimensional systems of three types. They must contain the following fragments: (1) conducting tubes with dielectric outside and inside them; (2) conducting tubes with dielectric between them; (3) stacks (tubes) of metal rings (clusters) with dielectric inside and outside them; (4) stacks (tubes) of metal rings (clusters) with dielectric between them. For the superconductivity emergence, these systems must have the required concentration and mobility of charge carriers, space (channels) for charge carriers flow, compression of the carriers flow and its focusing in the direction of spreading by the dielectric shell. The known low-temperature superconductors with a quasi-one-dimensional structure can serve as parents of such compounds.

### 3.1. Known inorganic superconductors with quasi-one-dimensional crystal structure

Let us consider the crystal structures of known superconductors and mark out their different quasi-one-dimensional fragments that could be responsible for superconductivity emergence.

*3.1.1. MRuX.* Superconducting transition temperatures of the compounds MRuX with the hexagonal $Fe_2P$-type structure are considerably high, about 13 K for h-ZrRuP and h-HfRuP [16, 17], 12 K for h-ZrRuAs [18, 19]. Usually, the structure of compounds of this type is considered as a layered one consisting of layers alternating along the *c* axis, occupied by Ru and X or M and X and separated by a distance of *c*/2. However, Seo et al. [20] and Hase [21], who performed studies of the electron structure of superconducting compounds h-ZrRuX (X = P, As, Si), relate superconducting properties to one-dimensionality of the electronic properties – partially filled one-dimensional band dispersive along the *c* direction. Hase demonstrates that the density of states at the Fermi energy is very sensitive to the in-plane Ru-Ru distance. Seo believes that this band is largely associated with the P(2)Ru3 prismatic chains along the *c* axis, whereas the shrinking of Ru-Ru bonds in triangular clusters does not have any role. The absence of the dependence of $T_c$ on just one parameter (Ru-Ru distance in clusters) was also indicated by experimental studies of h-MRuX [22-24].

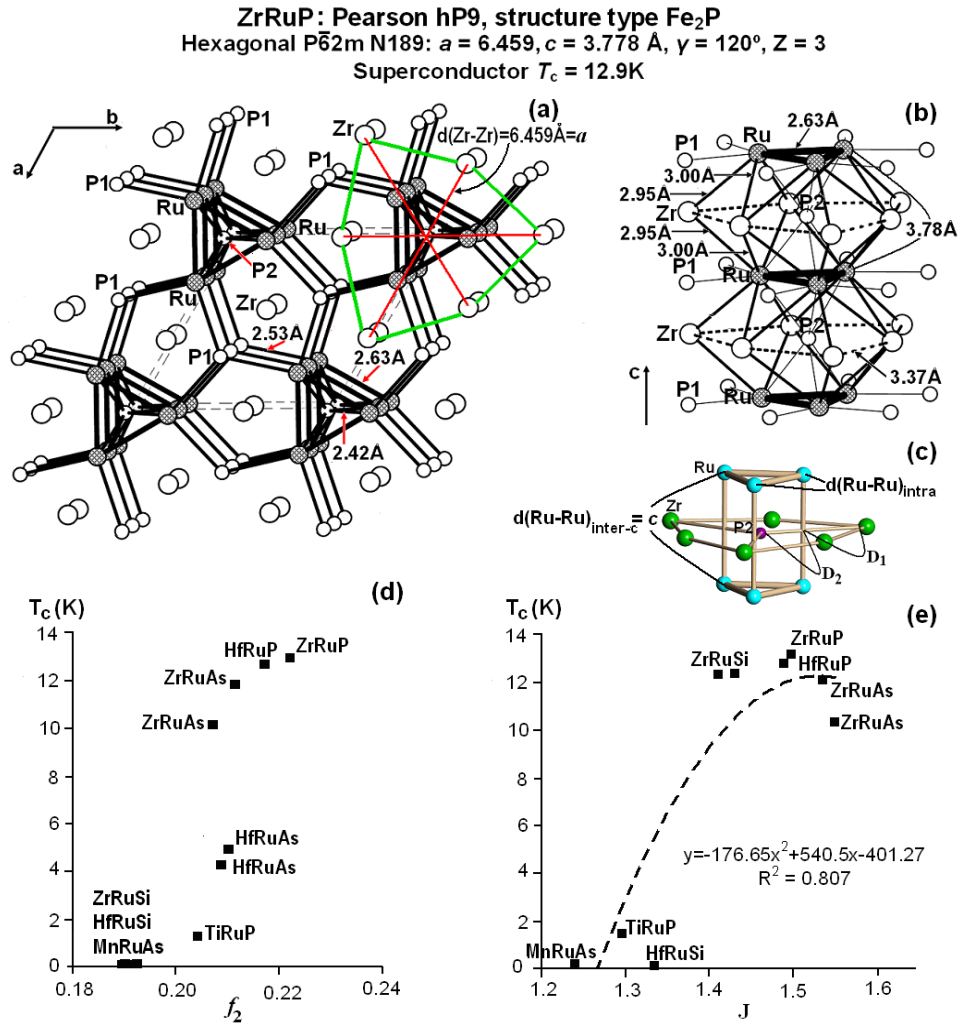

**Figure 4.** Crystal structure of ZrRuP [16] (a) quasi-one-dimensional fragment – stack of triangular metal clusters of $Ru_3$ with P2 atoms inside it and Zr and P1 atoms outside it (b) and structural parameters for calculation of correlations (c). $T_c$ as a function of $f_2$ ($f_2$ = d(Ru-Ru)$_{inter-c}$/ad(Ru-Ru)$_{intra}$) (d) and $J$ ($J$ = d(Ru-Ru)$_{inter-c}$/d(Ru-Ru)$_{intra}$($D_1$+$D_2$)) (e).

We will show quasi-one-dimensional features in the structure of h-MRuX that can be the basis to discuss $T_c$ in these compounds. For instance, let us consider the structure of ZrRuP [16] (figure 4(a) and (b)). The main fragment of this structure comprises stacks of two-dimensional triangular clusters of $Ru_3$ along the *c* axis. These stacks can be also considered as columns of trigonal prisms of $Ru_6$ centered by P2 atoms. The distances between Ru atoms in this quasi-one-dimensional fragment are the shortest ones in the structure (2.63 Å in cluster, 3.79 Å between clusters). These fragments are located in hexagonal cavities formed by triangular networks of Zr atoms and hexagonal networks of P1 atoms located in planes z = 1/2 and 0, respectively. The lengths of sides in the distorted Zr$_6$ hexagon and in the larger regular Si$_6$ hexagon are equal to 3.37 Å and 3.73 Å, respectively.

**Table 1.** Superconducting transition temperatures ($T_c$), lattice parameters, interatomic distances and dependence of $T_c$ on the ratio ($f_n$ and $J$) of these distances in ternary pnictides MRuX with the hexagonal Fe$_2$P-type ($P\bar{6}2m$, N189)

| Data for ICSD | Compound MRuX | $T_c$, K | $a$[a], Å | $c$[b], Å | d(Ru-Ru) intra[c], Å | $D_1+D_2$[d], Å | $f_1$ | $f_2$ | $J$[e] |
|---|---|---|---|---|---|---|---|---|---|
| - | ZrRuP [16] | 13 - 10.56 [16,17] | 6.459 | 3.778 | 2.629 | 0.959 | 1.437 | 0.222 | 1.498 |
| 53035 | HfRuP [26] | 12.70 - 11.08 [16] | 6.414 | 3.753 | 2.700 | 0.932 | 1.390 | 0.217 | 1.491 |
| 611301[f] | ZrRuAs [26] | 11.9 - 10.2 | 6.586 | 3.891 | 2.852 | 0.880 | 1.364 | 0.207 | 1.550 |
| 35593 | ZrRuAs [18] | [18,19] | 6.586 | 3.891 | 2.795 | 0.906 | 1.392 | 0.211 | 1.536 |
| 604605[f] | HfRuAs [18] | 4.93 - 4.37 | 6.568 | 3.842 | 2.787 | 0.896 | 1.378 | 0.210 | 1.538 |
| 610654[f] | HfRuAs [26, 27] | [18] | 6.568 | 3.842 | 2.799 | 0.876 | 1.372 | 0.209 | 1.566 |
| 77792 | TiRuP [26] | 1.33 [16] | 6.303 | 3.567 | 2.773 | 0.992 | 1.286 | 0.204 | 1.296 |
| 636584[f] | HfRuGe [28] | Not[g] [28] | 6.715 | 3.726 | 2.908 | 0.849 | 1.281 | 0.190 | 1.509 |
| 600432[f] | ZrRuSi [29] | 12.2 - 7 [23] | 6.660 | 3.655 | 2.861 | 0.895 | 1.278 | 0.192 | 1.429 |
| 16306 | ZrRuSi [30] | Not[g] [16] | 6.684 | 3.672 | 2.871 | 0.907 | 1.279 | 0.191 | 1.410 |
| 638841[f] | Hf Ru Si [31] | Not[g] [31] | 6.643 | 3.646 | 2.877 | 0.949 | 1.267 | 0.191 | 1.335 |
| 610918[f] | MnRuAs [32, 33] | Not[g] [34] | 6.518 | 3.619 | 2.879 | 1.013 | 1.257 | 0.193 | 1.241 |
| 650603[f] | ScRuSi [35] | Not[g] [31] | 6.851 | 3.423 | 2.967 | 0.954 | 1.154 | 0.168 | 1.210 |

[a] Distances between centers of Ru$_3$ clusters in the plane $ab$, equal to parameters $a$ and $b$.
[b] Ru-Ru distances between Ru$_3$ clusters along $c$-axis (d(Ru-Ru)$_{inter-c}$), equal to parameter $c$.
[c] Ru-Ru distances in Ru$_3$ clusters (d(Ru-Ru)$_{intra}$).
[d] ($D_1+D_2$) sum of effective distances between column edges and P and Zr atoms.
[e] J - ratio of crystal chemistry parameters (equation (6)).
[f] Atom positions and distribution estimated by the editor, not refined.
[g] Not superconducting above 1.2 K [1]

The characteristic feature of the crystal structure of MRuX compounds (table 1) consists in the presence of numerous bonds that are shorter than the sum of covalent radii [25], such as those between Ru atoms in triangular Ru$_3$ clusters (except the compound ScRuSi); between Ru atoms from Ru$_3$ clusters column and M atoms forming the cavity, in which this column is located (except the compounds TiRuP, MnRuAs, and ScRuSi); between M atoms in M$_6$ hexagons of this cavity (except the compounds TiRuP, MnRuAs, and ScRuSi), that indicates to the presence of metal bond. Also, the Ru$_3$ clusters are linked to each other through the Ru-X2 bond along the $c$ axis and Ru-X1 bonds in the basal plane. The bond lengths Ru-X1 and Ru-X2 in the compounds RuX are equal (d(Ru-P1) = 2.53 Å) or less (except d(Ru-Si1) = 2.62 Å in ScRuSi) than the sum of covalent radii by ~0.1 Å.

We investigated the dependence of $T_c$ of the compounds h-MRuX on the lengths of each of the above bonds and became convinced of its absence. However, there exists the dependence of $T_c$ on the ratio ($f$) of two main parameters characterizing quasi-one-dimensional fragments – stacks of two-dimensional triangular clusters of Ru$_3$: bond lengths in the cluster ($d$(Ru-Ru)$_{intra}$) and distances between clusters in stacks ($d$(Ru-Ru)$_{inter-c}$) equal to the parameter $c$ (table 1). This dependence can be written as:

$$f_1 = d(Ru-Ru)_{inter-c} / d(Ru-Ru)_{intra} \quad (3)$$

One should take into consideration that the structures of the compounds MRuX were not determined accurately, since the coordinates $x$ of M and Ru atoms were evaluated, in most cases, similarly to isotypic compounds and did not undergo refinement. The measurements of $T_c$ and structural parameters must have been performed on different samples, although superconductivity is known to be sensitive to the sample composition. For example, the samples ZrRuSi prepared by the arc-melted method did not manifest superconductivity above 1.2 K [16], whereas in the samples prepared at high temperatures and high pressures superconductivity emerges at around 7-12 K [23]. Here one can assume that we calculated the $f$ values for ZrRuSi (Table 1) from the structural data of samples that are not superconducting. The absence of a noticeable difference between the $f$ values for ZrRuAs (11.7 K) and HfRuAs (4.9 K), whose $T_c$ differ significantly, can be related to inaccuracy in evaluation of the coordinate $x$ Ru in HfRuAs that did not undergo correction.

In general, $T_c$ would depend not only on these two parameters characterizing quasi-one-dimensional fragments, but also on the distances between them (table 1). In the structure of $h$-MRuX, the distances between stacks can be considered as minimal ones ($d$(Ru-Ru)$_{inter-ab}$) between Ru atoms from adjacent stacks or as those between the centers of Ru$_3$ clusters in the plane $ab$ equal to parameters $a$ and $b$ ($a=b$). Moreover, the parameter $a$ is equal to the diameter of the outer shell from Zr atoms (d(Zr-Zr) = $a$), inside which Ru$_3$ columns are located (figure 4(a)).

These dependencies (figure 4(d)) can be written as:

$$f_2 = \frac{d(Ru-Ru)_{inter-c}}{d(Ru-Ru)_{intra}d(Zr-Zr)} \quad (4)$$

or

$$f_2 = c/d(Ru-Ru)_{intra} a \quad (5)$$

However, the best correlation with the transition temperature is provided by the ratio of d(Ru-Ru)$_{inter-c}$ to the product of the distance Ru-Ru inside this cluster (d(Ru-Ru)$_{intra}$) and the sum of effective distances (D$_1$+D$_2$) between column edges (Ru$_n$) and X and M atoms (figure 4(e)):

$$J = d(Ru-Ru)_{inter-c}/d(Ru-Ru)_{intra}(D_1 + D_2) \quad (6),$$

The effective distances were calculated with taken into account covalent radii (R) of M and X atoms and the number of electrons (z) in the valence shell:

$$D_1 = d(M-Ru_n) - R_M(z_M/4) \text{ and}$$

$$D_2 = d(X-Ru_n) - R_X(z_X/5) \quad (7)$$

where $z_M/4$ and $z_X/5$ are dimensionless coefficients. One should emphasize that the equations (6) and (7) are virtually identical to (1) and (2) for layered superconductors.

To sum up, the critical crystal chemistry parameters controlling the $T_c$ value in quasi-one-dimensional compounds MRuX are the Ru-Ru between Ru$_3$ clusters along triangular columns, the size of Ru$_3$ clusters, and effective distances between column edges and X atoms located inside the column and M atoms forming the outer shell.

The $T_c$ value is proportional to the distance between Ru$_3$ clusters in stacks d(Ru-Ru)$_{inter-c}$ (or parameter $c$) and inversely proportional to the distances between Ru atoms in clusters d(Ru-Ru)$_{intra}$ and the distance between stacks d(Ru-Ru)$_{inter-ab}$ (or parameter $a$). The presence of such dependence enables one to assume that the emergence of superconductivity in h-MRuX is related not to an individual cluster, but to a quasi-one-dimensional fragment – a stack of Ru$_3$ clusters along the axis $c$.

As seen from the presented correlations, the optimal characteristics of critical crystal chemistry parameters responsible for high $T_c$ in the compounds MRuX were already attained, and the possibilities of $T_c$ increase in these compounds were, as a result, exhausted. The reasons for low $T_c$ in the compounds of this type could consist in opposite effective charges of atoms inside (X) and outside (M) the tubes of the sandwich-type and overlapping of the transition space due to a strong covalent component of the bond between X atoms located inside the tube and Ru atoms. In our opinion, the same reasons are the main ones causing lower, than in superconducting cuprates, maximal possible transition temperatures in Fe-based pnictides. One more disadvantage in the structure of compounds MRuX consists in insufficient isolation of tubes of the sandwich-type from each other, since they are bonded by common M atoms (share M atoms) from the outer sandwich shell.

*3.1.2. Tl$_2$Mo$_6$Se$_6$.* In the crystal structure of the superconductor Tl$_2$Mo$_6$Se$_6$ [36] (figures 5(a) and (b)), triangular clusters of Mo$_3$ are about the same size ($d$(Mo-Mo) = 2.652 Å) as triangular Ru$_3$ clusters in the structure of h-ZrRuP – they forms stacks along the axis *c*. However, $T_c$ of $Tl_2Mo_6Se_6$ attains as low as 6.5 K [37]. Unlike the structure of h-ZrRuP, these stacks are so strongly compressed that the distances between $Mo_3$ clusters in the stack become equal to 2.241 Å. The increase of the internuclear repulsion upon compression forces triangular $Mo_3$ clusters to rotate relatively to each other by 180º (to get packed antiprismatically) and, as a result, attain acceptable distances (2.714 Å) between molybdenum atoms from adjacent clusters to preserve the system stability. In the end, quasi-one-dimensional clusters $(Mo_3)_\infty$ from facet sharing antiprisms (octahedra) $Mo_6$ are formed (figure 5(b)). They can be also considered as triangular screw tubes. All Mo-Mo distances in $(Mo_3)_\infty$ clusters are less than the sums of covalent radii of Mo (3.08 Å) that indicates to a strong metal bond. Se atoms remain outside these quasi-one-dimensional metal clusters – $(Mo_3)_\infty$ screw tubes and encapsulate them. They are bonded to Mo atoms by covalent bonds ($d$(Mo-Se) = 2.604 and 2.614 Å) forming chains of $(Mo_3Se_3)_\infty$ running along the hexagonal *c* axis separated by the Tl atoms. The Tl atoms bonds with chains are very weak ($d$(Tl-Se) = 3.393Å).

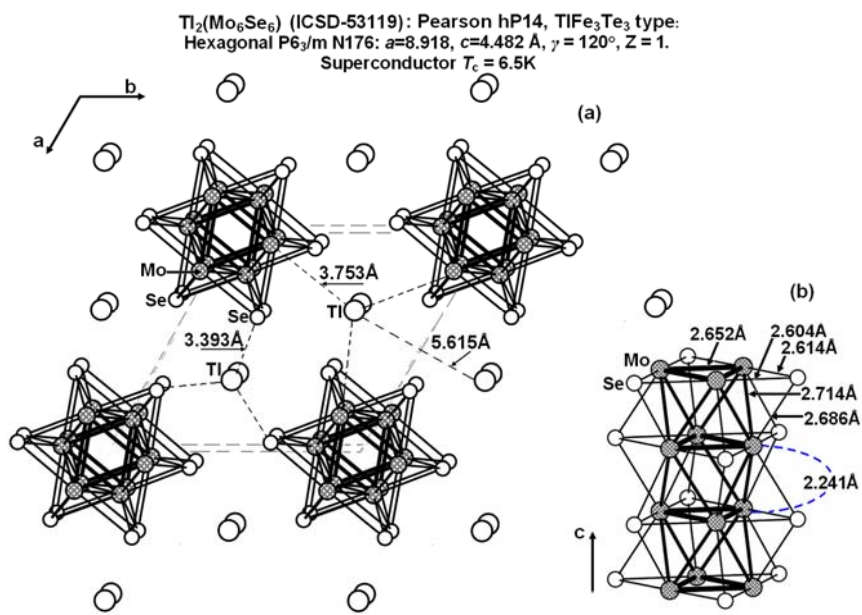

**Figure 5.** Crystal structure of $Tl_2Mo_6Se_6$ [36] (a) and quasi-one-dimensional fragment – triangular metal screw tubes $(Mo_3)_\infty$ with Se atoms outside them (b).

In the system $M_2Mo_6Se_6$ (*M* = Tl, In, Rb, Li, Na, K, Cs) [38], only $Tl_2Mo_6Se_6$ and $In_2Mo_6Se_6$ are superconducting, with $T_c$=3–6.6 K (varying between samples) and 2.9 K, respectively [39-41]. These two inorganic compounds with an expressed quasi-one-dimensional crystal structure are the most strictly anisotropic quasi-one-dimensional superconductors found until recently [42-45].

*3.1.3. $Lu_2Fe_3Si_5$.* According to [46-48], $Lu_2Fe_3Si_5$ is a weakly one-dimensional superconductor ($T_c$ = 6.1 K). The crystal structure of $Lu_2Fe_3Si_5$ [49] (figure 6(a)) contains two types of quasi-one-dimensional fragments along the axis *c*. The point is, which of these fragments could be responsible for the emergence of one-dimensional anisotropy? We tried to answer this question from the crystal chemistry point of view.

In the first fragment (figure 6(b)), the squares from Fe1 atoms are packed into stacks along the axis *c* at distances of 5.388 Å. Large-sized and rotated in the plane squares from Lu atoms are located in the middle between $Fe1_4$ squares. The distances between atoms in $Fe1_4$ ($d$(Fe1-Fe1) = 2.756 Å) and $Lu_4$ ($d$(Lu-Lu) = 3.620 Å) squares and between these squares ($d$(Fe1-Lu) = 3.040 Å) are less by 0.28 Å, 0.12 Å, and 0.15 Å, respectively, than the sum of the elements covalent radii ($R_{Fe}$ = 1.52 Å, $R_{Lu}$ = 1.87 Å [25]), which indicates to the presence of a metal bond. As a result, the stack of $Fe1_4$ and $Lu_4$ squares alternating along the axis *c* can be considered as a corrugated metal tube. Inside this tube, chains of Si3 atoms are located at alternating along the axis $c$ distances ($d$(Si3-Si3) = 2.68 and 2.71 Å) tha are much larger than the sum of covalent radii ($R_{Si}$ = 1.11 Å [25]). The Si3 atoms have covalent bonds with Fe1 atoms ($d$(Si3-Fe1) = 2.37 Å) and Lu atoms ($d$(Si3-Lu) = 2.90 Å). The silicon atoms Si1 ($d$(Si1-Fe1) = 2.37-2.38 Å and $d$(Si1- Lu) = 2.73-2.86 Å) and Si2 ($d$(Si2-Fe1) = 2.36 Å and d(Si2 - Lu) = 3.00-3.03 Å) covalently bonded to Fe and Lu atoms are located outside the tube as well.

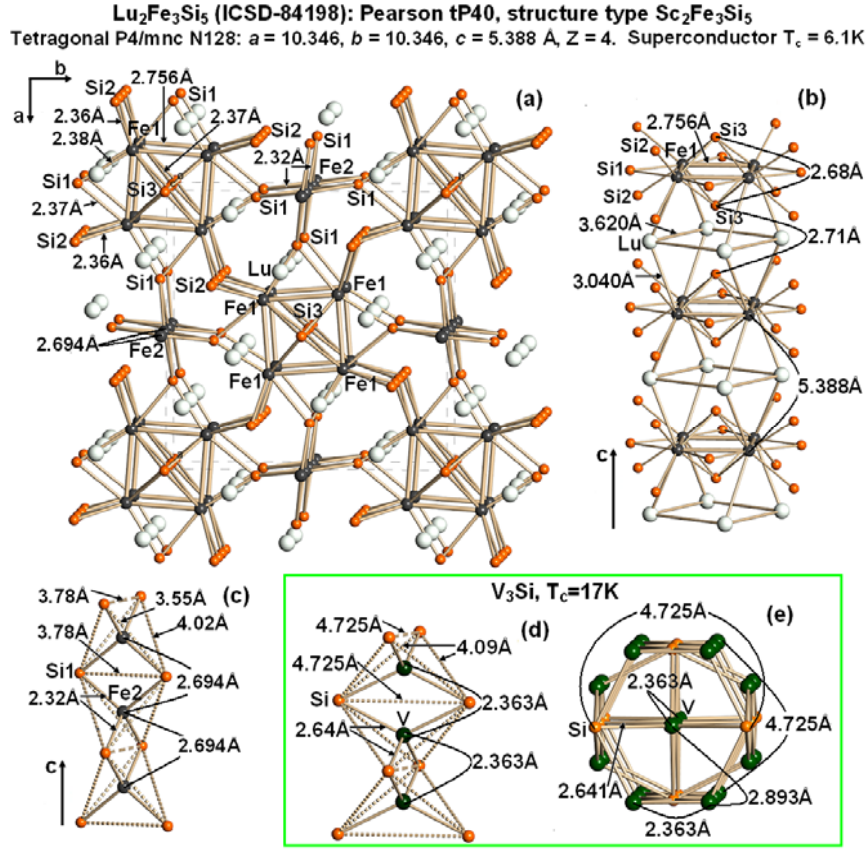

**Figure 6.** Crystal structure of $Lu_2Fe_3Si_5$ [49] (a) and quasi-one-dimensional fragments: (b) stack of metal clusters $Fe1_4$ with Si3 atoms chains inside it and Lu, Si1, and Si2 atoms outside it, (c) chain of $FeSi_4$ tetrahedra linked through common edges. Chains of $VSi_4$ tetrahedra linked through common edges (d) and channels with linear chains -V-V-V- (e) in the structure of $V_3Si$ [59].

**Table 2.** $T_c$ values and geometrical parameters of chains formed from tetrahedra linked through common edges in $A_3X$ superconductors (Cr3Si type (A-15), $cP8$, cubic $Pm\bar{3}n$) and in $Lu_2Fe_3Si_5$.

| ICSD | $A_3X$ | $T_c$, K [67] | $d(X$-$X)^a$, Å | $d(X$-$X) - 2R_X^b$, Å | $d(A$-$A)^c$, Å | $2R_A^b - d(A$-$A)^c$, Å |
|---|---|---|---|---|---|---|
| 76584 | $Nb_3Si$ [55] | *19.0* | 5.155 | 2.935 | 2.578 | 0.702 |
| 637221 | $Nb_3Ge$ [56] | 23.2 | 5.144 | 2.744 | 2.572 | 0.708 |
| 634745 | $Nb_3Ga$ [57] | 20.7 | 5.176 | 2.736 | 2.588 | 0.692 |
| 645502 | $Nb_3Sn$ [58] | 17.9 | 5.2887 | 2.509 | 2.644 | 0.636 |
| 652500 | $V_3Si$ [59] | 17.0 | 4.725 | 2.505 | 2.363 | 0.697 |
| 106076 | $Ta_3Sn$ [60] | 8.4 | 5.276 | 2.496 | 2.638 | 0.642 |
| 645347 | $Nb_3Sb$ [61] | *2.0* | 5.262 | 2.482 | 2.631 | 0.649 |
| 51996 | $Nb3In$ [62] | 9.2 | 5.293 | 2.453 | 2.647 | 0.633 |
| 105188 | $Nb_3Pb$ [63] | 8.0 | 5.319 | 2.399 | 2.659 | 0.621 |
| 603983 | $V_3Ge$ [64] | 6.2 | 4.781 | 2.381 | 2.391 | 0.669 |
| 58817 | $Nb_3Bi$ [65] | 3.1 | 5.320 | 2.360 | 2.660 | 0.620 |
| 106097 | $V_3Sn$ [66] | 3.8 | 4.975 | 2.195 | 2.488 | 0.572 |
| 84198 | $Lu_2Fe_3Si_5$ [49] | - | 3.780 | 1.560 | 2.694 | 0.346 |

[a] $d(X$-$X)$ – length of common edges of tetrahedral in the chain ($d(X$-$X) = a$ in structures of Cr3Si type).
[b] Covalent radii [25]: $R_{Si} = 1.11$, $R_{Ge} = 1.20$, $R_{Ga} = 1.22$, $R_{Sn} = 1.39$, $R_{Sb} = 1.39$, $R_{In} = 1.42$, $R_{Pb} = 1.48$, $R_{Pb} = 1.48$, $R_{Fe} = 1.52$, $R_V = 1.53$, $R_{Nb} = 1.64$ and $R_{Ta} = 1.70$Å.
[c] $d(A$-$A)$ – distance between metal atoms in the chain ($d(A$-$A) = a/2$ in structures of Cr3Si type).

So far, the role of Fe and Lu atoms in the emergence of superconductivity has remained partially unclear [50-54]. If just 3d–electrons of Fe are responsible for superconductivity, this superconducting fragment must be considered as a stack of metal clusters Fe1$_4$ with chains of Si3 atoms inside it and Lu, Si1, and Si2 atoms outside it. The reason of low temperature of transition ($T_c$) into Lu$_2$Fe$_3$Si$_5$ could consist in overlapping the space of carrier transfer inside the Fe1$_4$ clusters stack by electrons of the covalent bond Fe1-Si3 and the outside space by electrons of metal bonds Fe1-Lu and covalent bonds Fe1-Si1 and Fe1-Si2

The second one-dimensional fragment (figure 6(c)) in Lu$_2$Fe$_3$Si$_5$ comprises a chain of Fe$_2$Si$_4$ tetrahedra bonded through common edges. The Fe2 atoms in these fragments form linear metalchains along the axis $c$, where the distances between iron atoms ($d$(Fe2-Fe2) = 2.694) are less (by 0.35 Å) than the sum of covalent radii. Similar fragments built of AX$_4$ tetrahedra bonded through common edges exist in the structure of compounds A$_3$X with the structure A-15 (Fig. 6d), in which well-known superconductors V$_3$Si and Nb$_3$Sn, as well as Nb$_3$Ge having the maximal $T_c$ (23.2 K) for this type of compounds, crystallize. We decided to compare geometrical parameters of these one-dimensional fragments in order to understand whether the second fragment could be responsible for the emergence of a weakly one-dimensional superconductivity in Lu$_2$Fe$_3$Si$_5$. To solve this problem, it was necessary to find the correlation between $T_c$ and geometrical parameters of one-dimensional fragments in A$_3$X superconductors.

The characteristic feature of the structure A-15 consists in the fact that the transition atoms (A-atoms) form families of noncrossing linear chains that are stretched in three perpendicular directions and located in channels (figure 6(e)). The distances between A atoms in a single chain are significantly less than A-A distances to adjacent chains forming the channel walls. 'Free space' around linear metal chains emerged in A-15 superconductors as a result of a strong compression of AX$_4$ tetrahedra along one of twofold rotation axes. Such a compression yields: (1) decrease of the distances between A atoms in the chain and respective increase of the metal bond strength; (2) increase of the lengths of common edges of X-X tetrahedra and, as a result, channel widening; (3) increases of A-X distances and weakening of covalent bonding and concentration of electron density of binding electrons near the A atoms chain. As a result of studies of 12 A$_3$X superconductors (A = V, Nb and Ta; X = Si, Ge, Sn, Pb, Ga, In, Sb, and Bi) (table 2), we revealed that $T_c$ increases along with the increase of the difference between the length of the tetrahedra common edge $d$(X-X) and the sum of covalent radii of X. Just two compounds (Nb$_3$Si and Nb$_3$Sb) manifest a significant deviation from the $T_c(d(X\text{-}X) - 2R_X)$ correlation, according to which the transition temperature $T_c$ must be significantly higher than 23 K and 8 K, respectively. Predictions of high $T_c$ (in the range 25-38 K) of the stoichiometric compound Nb$_3$Si were made on the basis of different arguments in other works as well [67-70]. Aside from this dependence, for each individual family of Nb$_3$X and V$_3$X one observes the dependence of $T_c$ on distances between A atoms in the chain. The temperature $T_c$ increases along with the decrease of the A-A distance. Thus, we established that $T_c$ of A$_3$X superconductors correlates to the cross-section size of the channel, in which the A atoms chain is located, and the distances between atoms in this chain, i.e., the strength of metal bond in it. As was shown above, the possibility of emergence of superconductivity and the $T_c$ value in layered materials are determined by the following critical parameters: concentration of charge carriers, distances between charge carrier atoms in metal layer, and parameters of the cross-section size of the space between metal layer and dielectric layers from both sides.

The critical parameters of chains from tetrahedra bonded with common edges differ substantially in Lu$_2$Fe$_3$Si$_5$ and V$_3$Si ($T_c$ = 17 K) (figure 6(e), table 2), in spite of almost identical covalent radii of Fe ($R_{Fe}$ = 1.52 Å) and V ($R_V$ = 1.53 Å) metals. In Lu$_2$Fe$_3$Si$_5$ the distances between iron atoms in the chain are larger by

12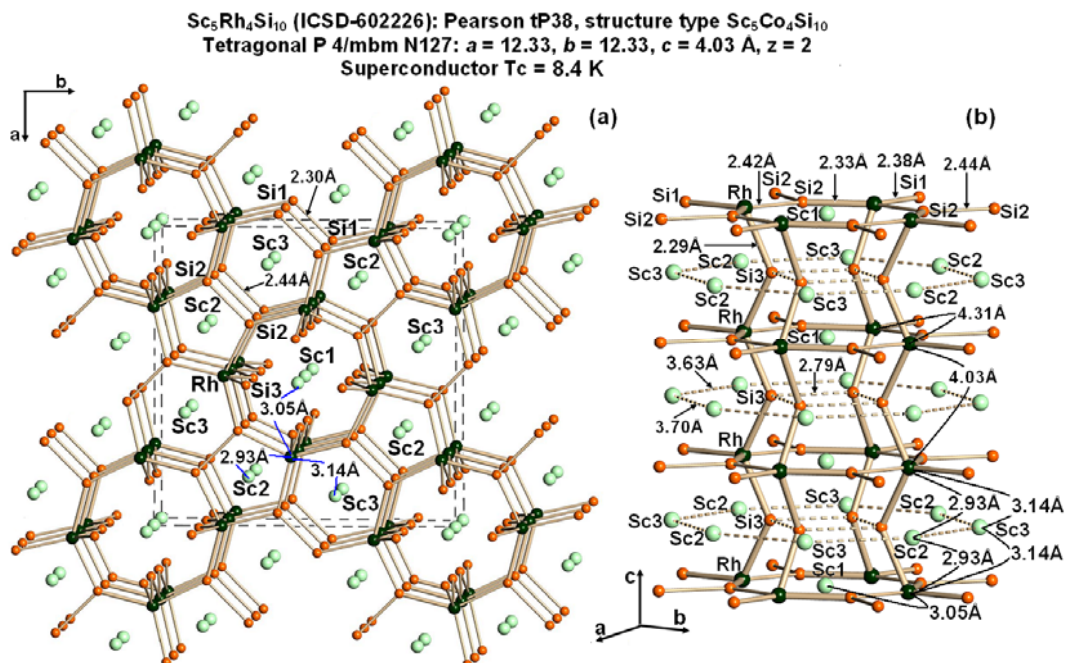

**Figure 7.** Crystal structure of $Sc_5Rh_4Si_{10}$ (a) and quasi-one-dimensional fragment – corrugated octagonal tube from covalently bonded Rh and Si atoms with Sc atoms inside and outside it (b).

0.33 Å, whereas the length of tetrahedra common edges are, conversely, smaller by 0.95 Å than in $V_3Si$. For the emergence of superconductivity based on the above chains in $Lu_2Fe_3Si_5$, at least with $T_c$ ~3 K, the chains must be compressed until the state when the distance between iron atoms decreases down to 2.45 Å, whereas the length of common edges increases up to 4.42 Å.

So it can be concluded that the second fragment cannot be superconducting, whereas the emergence of a weak one-dimensional superconductivity in $Lu_2Fe_3Si_5$ must be related to the first quasi-one-dimensional fragment comprising a stack of metal clusters $Fe1_4$ with Si3 atoms chains inside it and Lu, Si1, and Si2 atoms outside it.

*3.1.4. $Sc_5Rh_4Si_{10}$.* According to [71, 72, 47, 73], ternary silicides and germanides $R_5T_4X_{10}$ with the structure of $Sc_5Co_4Si_{10}$, where $R$ = Sc, Y, and rare-earth elements, $T$ = Co, Rh, Ir, or Os, and $X$ = Si, Ge [74-76], are quasi-one-dimensional superconductors. Among $R_5T_4X_{10}$ compounds, $Sc_5Rh_4Si_{10}$ (figure 7), $Sc_5Ir_4Si_{10}$, and $Y_5Os_4Ge_{10}$ manifest superconductivity with the highest $T_c$ of 8.4, 8.6, and 9.10 K, respectively [77]. To explain one-dimensional anisotropy in the superconducting properties, one marks out in the crystal structure of these compounds different one-dimensional fragments along the axis $c$, for example, zigzag chain T-X-T atoms or chains of R1 linked by short bonds with X3 atoms. Sometimes one marks out more complex formations, for example, in the form of 1D chains of R1 atoms (bonded to T and X3 atoms), which are embedded in a network of closely bonded R2 and R3 atoms [78]. However, the chains of R1 atoms along the axis $c$ have only an indirect relation to the one-dimensional character of this structure due to the following reasons: i) strength of the covalent bond between $R$1 and $X$3 atoms in this chain is not very high; ii) metal bond between $R$1 and $T$ is realized only in the plane $ab$ and is not strong as well; iii) there is no metal bond between $R$1 atoms along the axis $c$ and between R2 and R3 atoms.

In structural terms, in these silicides one can mark out another quasi-one-dimensional fragment along the axis $c$ based on transition $d$-elements. Let us demonstrate this on the example of the compound $Sc_5Rh_4Si_{10}$ (figure 7(a)). The main structural unit of this compound is a square pyramid $RhSi_5$, in which the bonds Rh-Si (2.29-2.42 Å) are covalent. The length of these bonds is far smaller than the sum of covalent radii (2.53 Å). $RhSi_5$ pyramids are linked by four units through four Si2 atoms located in the pyramids bases and form octagonal rings $Rh_4Si_4$ in the plane $ab$. The lengths of Rh-Si2 bonds in the ring equal to 2.33 and 2.42 Å are in alternating order.

In their turn, these four-pyramid units are linked to each other by Si3 atoms (d(Rh-Si3) = 2.29 Å), also located in the pyramid bases, into corrugated tubes (figure 7b) stretched along the axis *c*.

The Sc1 atoms are located inside these tubes in the centers of $Rh_4Si_4$ rings. They are coupled by metal bonds with four Rh atoms from the $Rh_4Si_4$ ring and by covalent bonds with eight Si3 atoms. The metal bond Sc1-Rh (3.045 Å) is not strong, since its length is just a bit smaller than the sum of covalent radii (3.12 Å) of these atoms. The length of the bond of Sc1 atom with Si3 atoms ($d$(Sc1-Si3) = 2.82 Å) linking them into a chain along the axis *c* slightly exceeds the sum of covalent radii (2.81 Å) of these atoms. There is no metal bond between Sc1 atoms along the axis *c*, since the distances Sc1-Sc1 (4.030 Å) exceed substantially (by 0.63 Å) the sum of Sc covalent radii (3.40 Å).

The vertices of square pyramids (Si1 atoms) are directed outside from tubes and link them to each other. Extra bonds between tubes emerge due to approaching of Si1 ($d$(Si1-Si1) = 2.30 Å) and Si2 ($d$(Si2-Si2) = 2.44 Å) atoms of adjacent tubes at distances slightly exceeding the sum of covalent radii of Si (2.22 Å). As a result, there emerge pentagonal and hexagonal channels from Rh and Si atoms filled by Sc2 and Sc3 atoms, respectively. The Sc2 and Sc3 atoms form flat octagonal rings of larger size ($d$(Sc2-Sc3) = 3.632 Å and 3.697 Å) than that of $Rh_4Si_4$, between which the former are located. There is no metal bond between Sc2 and Sc3 atoms in these rings, but it is present with Rh atoms ($d$(Sc2-Rh) = 2.927 Å, $d$(Sc3-Rh) = 3.045 Å) from $Rh_4Si_4$ rings.

To sum up, the corrugated octagonal tubes built from covalently bonded Rh and Si atoms, inside and outside which Sc atoms are located (figure 7(b)), can be responsible for the emergence of quasi-one-dimensional superconductivity in $Sc_5Rh_4Si_{10}$. The fact that the distances between Sc atoms and Rh atoms from tubes are smaller (by 0.19 and 0.08 Å) than the sum of covalent radii is similar to the superconducting sandwich in $YBa_2Cu_3O_7$ [79], where the distances ($d$(Cu2-Y) = 3.206 Å) between Y atoms and Cu atoms from the $CuO_2$ layer are smaller (by 0.014 Å) than the sum of covalent radii of Y and Cu (3.22 Å). The quasi-one-dimensional fragment in $Sc_5Rh_4Si_{10}$ we marked out has some similarities with the first fragment in $Lu_2Fe_3Si_5$ (figure 7(b), but it has no metal bonds between atoms of *d*-element (Rh).

*3.1.5. $Yb_7Co_4InGe_{12}$.* In quaternary compounds $R_7Co_4InGe_{12}$ (R = Dy, Ho, Yb) [80] (figure 8) quasi-one-dimensional fragments in the form of corrugated octagonal tubes from cobalt and germanium atoms, inside and outside which R atoms are located, are similar to the fragments in structures of the $Sc_5Co_4Si_{10}$ type. However, these compounds are not superconductors. If one substitutes Yb by Y and Co by Os in the compound $Yb_7Co_4InGe_{12}$, it is possible that the compound $Y_7Os_4InGe_{12}$ would become a superconductor, just like $Y_5Os_4Ge_{10}$.

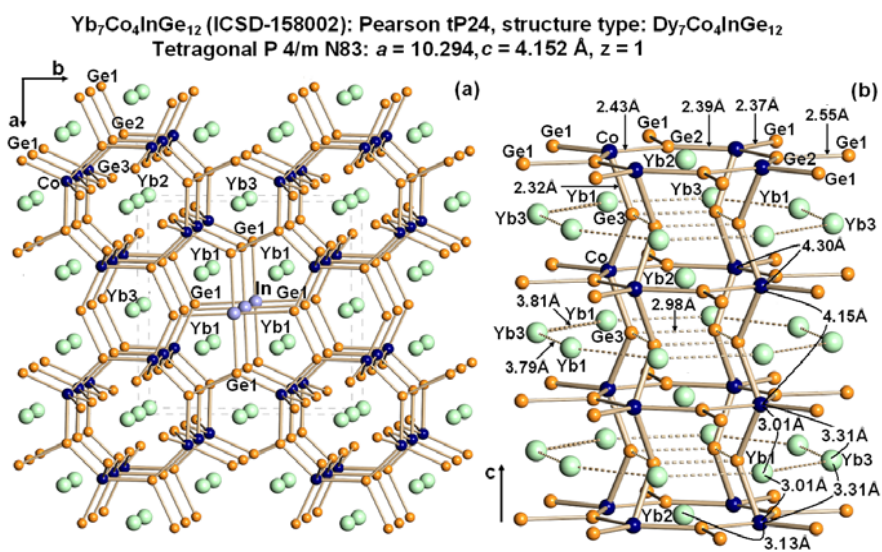

**Figure 8.** Crystal structure of $Yb_7Co_4InGe_{12}$ [80] (a) and quasi-one-dimensional fragment – corrugated octagonal tube from cobalt and germanium atoms with Yb atoms inside and outside it Yb (b).

## 3.2. Models of superconductors with quasi-one-dimensional crystal structure

In the inorganic crystal structure database (ICSD) we identified 5 compounds ($La_4BaCu_4FeO_{12.96}$ (ICSD-96335), $Ba_9Cu_7O_{15}Cl_2$ (ICSD-9628), KCuO (ICSD-40158), $LaGaBi_2$ (ICSD-165559) and $NaBa_6Cu_3Te_{14}$ (ICSD-82671)), whose structures generally comply with the crystal chemistry parameters we revealed, at which the emergence of superconductivity with high temperatures of superconducting transition is possible. These compounds can be parents of new families of inorganic superconductors with quasi-one-dimensional crystal structure, if one performs respective substitutions and doping with holes.

### 3.2.1. Oxygen-deficient perovskites $La_4BaCu_5O_{13}$ and $La_4BaCu_4FeO_{12.96}$.
Since the discovery of copper oxide superconductors [4], dozens of works devoted to findings of superconductivity at a temperature 165K and higher, up to 300 K, in new layered copper oxide systems with structures of the perovskite type have been reported. As a rule, the reports contained the information that not the whole sample, but just its small part manifests superconducting properties at such high temperature. However, in all cases these results were not reproducible and, most probably, inaccurate. In view of this, we posed the question whether this superconducting part could be a copper oxide system with the quasi-one-dimensional structure. We have analyzed the crystal structures of copper oxide systems available in the *ICSD* database and found the compound $La_4BaCu_4FeO_{12.96}$ [81], which is suitable, according to our criteria, to be the model of superconductors with a quasi-one-dimensional structure after appropriate substitutions. Interestingly, this compound belongs to the very system of oxygen-deficient perovskites $La_4BaCu_5O_{13+\delta}$ [82], whose study with respect to metal conductivity led Bednorz and Müller to search of high-temperature superconductivity in Ba-La-Cu oxides [83].

$La_4BaCu_4FeO_{12.96}$ has been reported by Anderson et al. [81] to crystallize in a tetragonal structure with *P4/m* space group ($a$ = 8.638, $c$ = 3.904 Å, z = 1). According to the conventional description of the crystal structure of perovskite $La_4BaCu_5O_{13+\delta}$ [81, 82, 84], the structure of $La_4BaCu_4FeO_{12.96}$ (figure 9(a)) has groups of four corner-sharing $CuO_5$ pyramids linked through $FeO_6$ octahedra. Each octahedron shares four corners with four pyramids and two corners with two other octahedra; each pyramid is linked to four other pyramids and one octahedron. The framework exhibits one perovskite-like tunnel and two hexagonal tunnels per cell. $Ba^{2+}$ and $La^{3+}$ are ordered with 12-coordinate $Ba^{2+}$ in perovskite tunnels and 10-coordinate $La^{3+}$ in hexagonal tunnels. $La_4BaCu_4FeO_{12.96}$ is an example of a successful occupation of the octahedral site in the $La_4BaCu_5O_{13+\delta}$ phase by Fe1(*a*) ions, and on its basis it is easier to observe quasi-one-dimensional features of the structural type $La_4BaCu_5O_{13}$ and represent the structure of these compounds in a different way by stressing their inherent quasi-one-dimensional anisotropy.

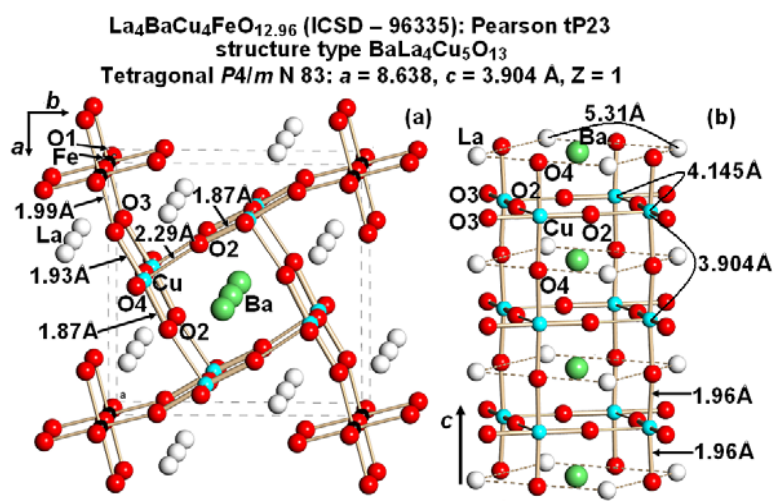

**Figure 9.** Crystal structure of $La_4BaCu_4FeO_{12.96}$ [81] (a) and quasi-one-dimensional fragment – tetragonal copper-oxygen tube with Ba ions inside it and La ions outside it (b). Axial and basal Cu-O bonds of $CuO_5$ pyramids alternate along square sides in the plane *ab*.

The main difference of a structure of the type $La_4BaCu_5O_{13}$ from that of layered high-$T_c$ cuprates consists in the fact that the former perovskite structure is divided in the plane *ab* into individual perovskite-type units shaped into columns along the axis *c* rather than perovskite-type layers perpendicular to the same axis *c*. From this point of view, the structure of $La_4BaCu_4FeO_{12.96}$ consists of two types of one-dimensional fragments stretched along the axis *c*: tetragonal copper-oxygen tubes (figure 9(a) and (b)) built from networks of distorted copper-oxygen squares and linear chains of top-linked $FeO_6$ octaherda. Tubes and chains are linked to each other through common oxygen atoms O3. Ba ions are located inside the tubes, whereas La ions are located outside them in the formed cavities. In tubes the Cu-O2 bond lengths in the plane *ab* are unequal (1.87 and 2.29 Å) and alternate along the square sides, whereas along the axis *c* the Cu-O3 bonds lengths are identical and equal to 1.96 Å. The latter is concerned with the fact that square pyramids $CuO_5$ having a characteristic Jahn–Teller distortion are linked in the plane *ab* alternately through O2 oxygen atoms located in the pyramids vertices and bases and along the axis *c* just by oxygen atoms located in pyramids bases. Anisotropy of the crystal structure of compounds of this type corresponds to that of the electrical transport property in $La_4BaCu_5O_{13}$ [85]: in the *a-b* plane, it is semiconducting, while along the *c*-direction, it is metallic.

$La_4BaCu_5O_{13+\delta}$ and $La_4BaCu_4FeO_{12.96}$ are known to be non-superconducting [81, 86]. We believe that attaining the optimal concentration of charge carriers in copper-oxygen tubes could yield the emergence of the superconducting state. Variation of the charge carriers concentration through compression or stretching of Cu-O bonds can be realized by heterovalent and isovalent substitution of Cu1(Fe), La, and Ba ions occupying the 1(*a*), 4(*k*), and 1(*d*) positions, respectively. The problem of substitutions consists in preserving the superconducting fragment comprising quasi-one-dimensional copper-oxygen tubes with positively charged ions inside and outside them and preventing the increase of the oxygen content above 13 leaving the following positions occupied by oxygen atoms: O1 – 1(*b*), O2 – 4(*j*), O3 – 4(*j*), and O4 – 4(*k*). Deficit of the oxygen atoms O1 (position 1(*b*)) binding (Cu1(Fe)) octahedra into a chain would enable one to additionally adjust the carriers concentration and would not cause the destruction of this structural type. Otzschi et al. [87] managed to obtain the compound $(La_{0.8333}Sr_{0.1666})_5Cu_5O_{13}$, whose structure was similar to $La_4BaCu_5O_{13}$, but lanthanum and strontium in the structure were statistically distributed, in contrast to $La_4BaCu_5O_{13+\delta}$, where lanthanum and barium were located in an ordered manner. Comparison of the Cu2-O bond lengths in copper-oxygen tubes of the compounds $La_4BaCu_4FeO_{12.96}$ and $(La_{0.8333}Sr_{0.1666})_5Cu_5O_{13}$ demonstrates the expected results. Substitution of a large barium ion located inside the tube by a mixture of smaller La and Sr ions resulted in shortening of Cu2-O2 bonds in the plane *ab* by 0.02 – 0.04 Å Cu2-O4 bonds along the axis *c* by 0.04 Å. Substitution of a part of La ions located outside the tube by larger Sr ions increased the length of the bonds of the tube with the octahedra chain (Cu2-O3) by 0.075 Å.

Aside from the variation of the carriers' concentration, we suggest an attempt to slightly modify the structural type $La_4BaCu_5O_{13}$. The purpose of this modification consists in strengthening of one-dimensional anisotropy – 'isolation' of copper-oxygen tubes and achievement of similarity of the structure of copper-oxygen tubes with $CuO_2$ layers. For this purpose, it is necessary just to rotate the pyramids $Cu2O_5$ in the tube with apical vertices looking outside and significantly increase the distance to these vertices (figure 10(a)). Then the copper-oxygen tube (figure 10(b)), just like the $CuO_2$ layer in high-$T_c$ cuprates, will be formed from vertex-linked bases of $CuO_5$ pyramids, whose apical vertices would be directed outside, and the Cu-O bonds in the tube will be much shorter than the bonds to apical vertices. In fact, one needs to strongly compress the tube by substituting large Ba ions located inside it by small ones (Ca, Mg, Y, La), whereas small La ions located inside, in opposite, must be substituted by larger ions (Ba or Sr). Here both positions must contain, along with two-valence ions, three-valence ones (La, Y, or Tl) to preserve the charge compensation. Copper ions in $Cu1O_6$ octahedra can be left or substituted by $Hg^{2+}$, $Tl^{3+}$, or $Pb^{2+,4+}$. All the above substitutions were successfully performed in

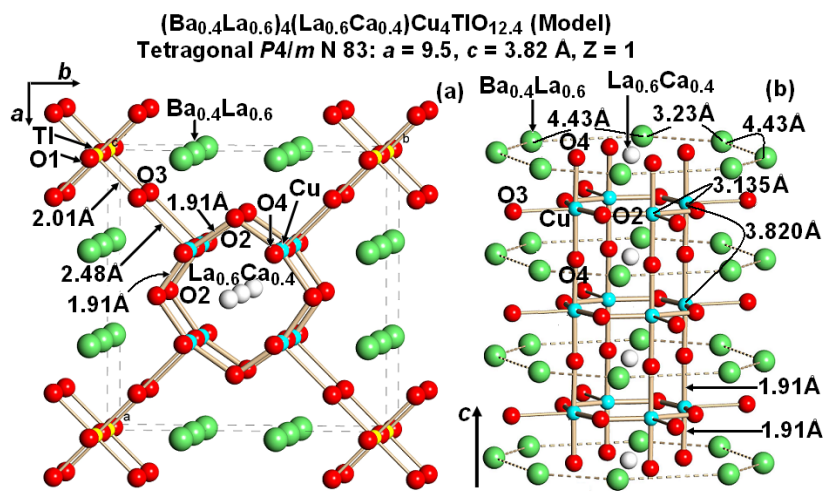

**Figure 10.** Modification of the structural type $La_4BaCu_5O_{13}$: crystal structure $(Ba_{0.4}La_{0.6})_4(La_{0.6}Ca_{0.4})Cu_4TlO_{12.4}$ (model) (a) and quasi-one-dimensional fragment – copper-oxygen from $CuO_5$ pyramids, whose axial vertices are directed outside the tube (b). Small cations $(La_{0.6}Ca_{0.4})$ are located inside the tube, large cations $(Ba_{0.4}La_{0.6})$ are located outside the tube.

high-$T_c$ cuprates [see, for example, 88-90]. The composition of these hypothetical superconductors can be as follows: $(Ba_{0.4}A_{0.6})_4(A_{0.6}Ca_{0.4})Cu_4BO_{13-\delta}$ (where A = $Tl^{3+}$, La, Y; B = $Tl^{3+}$, Cu, $(Hg_{1-x}Tl_x)$, $(Hg_{1-x}Cu_x)$) and $(Sr_{0.4}Y_{0.6})_4(Y_{0.6}Ca_{0.4})Cu_4BO_{13-\delta}$ (where B = Cu, $(Tl_{1-x}Pb_x)$). We represented their structure in the tetragonal space group P4/mmm. For the $(Ba_{0.4}La_{0.6})_4(La_{0.6}Ca_{0.4})Cu_4TlO_{12.4}$ structure (figure 10) the lattice parameters are $a$ = 9.5, $c$ = 3.82 Å, z = 1 with atomic positions of $(Ba_{0.4}La_{0.6})$ at 4$m$ (0.33, 0, 1/2), $(La_{0.6}Ca_{0.4})$ at 1$d$ (1/2, 1/2, 1/2), Cu at 4$j$ (0.335, 0.335, 0), Tl at 1$a$ (0, 0, 0), O1 at 1$b$ (0, 0, 1/2), O2 at 4$n$ (0.22, 1/2, 0), O3 at 4$j$ (0.15, 0.15, 0) and O4 at 4$k$ (0.337, 0.337, 1/2). One should mention that there exist oxygen vacancies in the position O1 (position occupancy q = 0.4). The Cu-O distances in the $CuO_5$ pyramid are equal to 2.485 Å, 1.911 Å, and 1.910 Å to apical vertex (O3), two basal vertices (O2) in the plane $ab$, and two basal vertices (O4) along the axis $c$, respectively. The number of charge carriers per copper atom is equal to 0.25 according to evaluation made from the copper bond valence sum (BVS) [11].

Thus, oxygen-deficient perovskites of the structural type $La_4BaCu_5O_{13}$ contain quasi-one-dimensional fragments – conducting copper-oxygen tubes with dielectric (positively charged ions) inside and outside them. In essence, these fragments are similar to the Ginzburg sandwich (*ISI*). Besides, one should emphasize similarity of the structures of these fragments to those in low-temperature superconductor $Sc_5Rh_4Si_{10}$ (see above) comprising rhodium-silicon tubes with Sc atoms located inside and outside them. We assume that these compounds could be the candidates yielding superconductors upon respective substitutions or structural modifications

*3.2.2. $Ba_9Cu_7O_{15}Cl_2$.* Among oxocuprates and halooxocuprates, the structure of $Ba_9Cu_7O_{15}Cl_2$ is a unique one. $Ba_9Cu_7O_{15}Cl_2$ [91, 92] (figure 11(a)) crystallizes in hexagonal space group *P6/mmm* ($a$ = 11.257 and $c$ = 5.853 Å, Z = 1). Copper ions occupy two crystallographically independent sites Cu1 (1$b$) and Cu2 (6$l$) differing by the surrounding coordination. Cu2 has the well-known square $O^{2-}$ coordination. Flat square $CuO_4$ fragments are linked by edges and form $Cu_6O_{12}$ ring clusters. Within the ring clusters, the Cu2-O2 distances are 1.899 Å, the Cu2-O2-Cu2, bond angles are 84.5°, and the Cu2-Cu2 distances (2.554 Å) are less than the sum of covalent radii (2.64 Å), which indicates to the presence of metal bond in $Cu_6$ rings. One should mention that the Cu2-O2 distance in $CuO_4$ squares of this compound and the number of holes (p = 0.21) in $Cu_6O_{12}$ ring clusters are close, according to estimations on copper BVS, to respective values (1.891 Å and p = 0.26) in the internal plane of $HgBa_2Ca_2Cu_3O_{8.16}$ [93] responsible for the emergence of superconductivity at $T_c$ = 145.5 under pressure of 6.0 GPa. Similar $Cu_6O_{12}$ clusters, but of a larger size ($d$(Cu-O) = 1.96 Å, Cu-O-Cu = 82.9°, and (Cu-Cu) = 2.605 Å), can be observed in another compound ($BaCuO_2$) [94-97]; however, 8 of such clusters in a large cubic unit cell (lattice parameter $a$ = 18.28 Å) are distributed in a different fashion and do not form one-dimensional fragments.

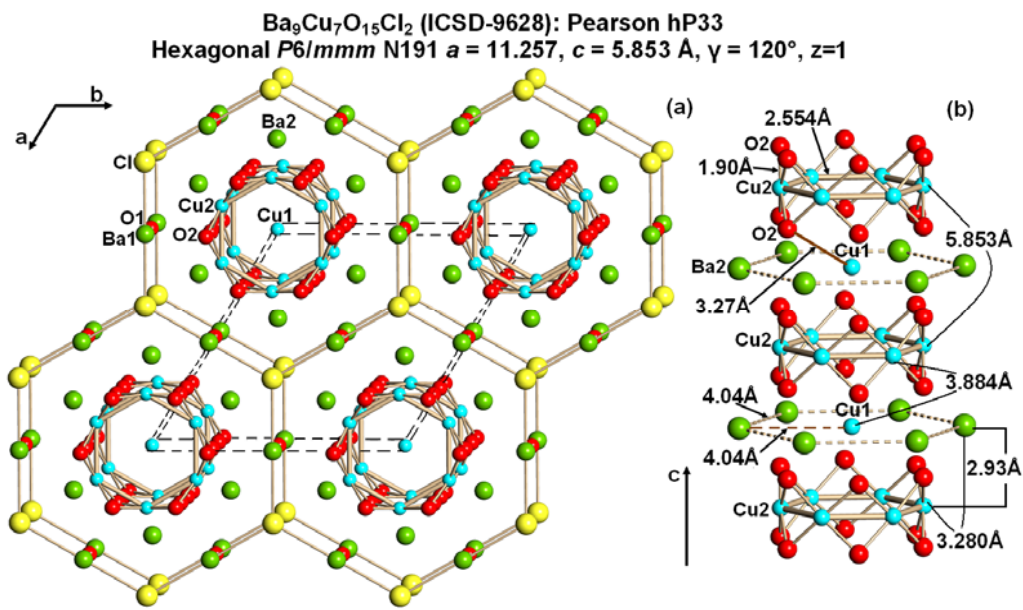

**Figure 11.** Crystal structure of $Ba_9Cu_7O_{15}Cl_2$ [91] (a) and quasi-one-dimensional fragment – hexagonal tube from $Cu_6O_{12}$ ring clusters with Cu1 ions inside it and Ba2 ions outside it (b).

The main feature of the structure of $Ba_9Cu_7O_{15}Cl_2$ consists in the fact that $Cu_6O_{12}$ rings are packed in stacks along the axis $c$ and comprise quasi-one-dimensional fragments – hexagonal tubes (figure 11(a)). The diameter of $Cu_6$ rings is equal to 5.108 Å, whereas the Cu2-Cu2 distances between rings in the stack and between stacks are equal to 5.853 Å and 6.833 Å, respectively. Large cavities between ring clusters are occupied by Cu1 atoms ($d$(Cu1-Cu1) = 5.853 Å and $d$(Cu1-Cu2) = 3.884 Å). As a result of such positioning, copper atoms acquire an uncharacteristic biplanar 12-fold coordination ($d$(Cu1-O2) = 3.266 Å). According to the BVS estimation, the charge of Cu1 atoms is as low as 0.16. Tubs of $Cu_6O_{12}$ rings are centered by Cu1 atoms, surrounded from the outside by hexagonal rings of Ba2 atoms ($d$(Ba2-Ba2) = 4.036 Å, $d$(Ba2-Cu2) = 3.884 Å and $d$(Ba2-O2) = 2.698 Å), and placed into large hexagonal channels forming linear -Ba1-O1- chains ($d$(Ba1-O1) = 2.927 Å), bonded to each other by Cl ions ($d$(Ba1-Cl) = 3.249 Å), along the axis $c$.

To realize the transition of $Ba_9Cu_7O_{15}Cl_2$ into the superconducting state, one can vary the number of carriers and the values of important interatom distances, for example, by hole-doping of $Cu_6O_{12}$ rings through introduction of Na or K atoms into the positions Ba1 and Ba2 and Li, Na, or K atoms into the position Cu1. Through isovalent substitutions of large ions $Ba^{2+}$ and $Cl^-$ by smaller ones $Sr^{2+}$, $Ca^{2+}$, and $F^-$, one can reduce the Cu2-Cu2 distances between clusters inside tubes (along the axis $c$) and between them (in the plane $ab$). To attain similarity to high-$T_c$ cuprates, one can substitute atoms of Cu1 by those of Ca or Y. The latter would not yield a substantial decrease of the holes number, since the elementary unit contains only one Cu1 atom and Cu1-O distances are very big. It is possible to attempt substituting Cl atoms by those of oxygen, but this must result in structural changes.

We assume that the quasi-one-dimensional fragment comprising a tube of metal rings $Cu_6O_{12}$ with a dielectric cation chain inside the tube and surrounded by dielectric cation ring outside it could cause the emergence of superconducting properties in this compound. The principal difference of this quasi-one-dimensional fragment from the Ginzburg sandwich consists in the fact that in the former the tube with charge carriers is not continuous, as the $CuO_2$ layer in the sandwich, but consists of metal clusters. However, this fact does not hamper their transition into the superconducting state, since there are well-known superconductors (for example, $Lu_2Fe_3Si_5$ and RuZrP, see above) with the quasi-one-dimensional cluster structure, but low transition temperature. Nevertheless, one has grounds to expect high $T_c$ in compounds of the structural type $Ba_9Cu_7O_{15}Cl_2$. The point here is in the fact that, unlike superconductors $Lu_2Fe_3Si_5$ and RuZrP, in the quasi-one-dimensional fragment $Ba_9Cu_7O_{15}Cl_2$ there is virtually no overlapping of the space of charge carriers' transfer by binding electrons along the tube from $Cu_6O_{12}$ rings, just like in the internal $CuO_2$ of plane high-$T_c$ cuprates with the number of $CuO_2$ planes equal to 3.

*3.2.3. KCuO and RbCuO.* Binary oxides ACuO (A = Li, Na, K and Rb) [98-102] crystallize in a tetragonal system, but their space group was not immediately established unambiguously. First, it was believed that the space group of these compounds was $I\bar{4}$ N82 [98, 99], then $I\bar{4}m2$ N119 [100, 101] while later [102] the conclusion was put forward that for all the compounds (except, possibly, RbCuO) the centrosymmetric space group I4/mmm N139 was the most relevant. It is possibly related to the presence of vacancies in A cations positions.

The structure of ACuO oxides (figure 12(a)) contains two types of quasi-one-dimensional fragments stretched along the axis *c*: stacks of square rings $[Cu_4O_4]^{4-}$ (figure 12(b)) and chains of common edge-bonded $A_4^{4+}$ tetrahedra located between stacks. Tetrahedra chains are linked to oxygen atoms from $[Cu_4O_4]^{4-}$ rings. The ACuO structures in I4/mmm and $I\bar{4}m2$ space groups differ only by the fact that in the first $[Cu_4O_4]^{4-}$ rings are flat, whereas in the second one oxygen atoms slightly deviate from $Cu_4$ square plane.

Further we will show changes of the most important structural parameters in a row A = Li, Na, K and Rb within the space group *I*4/*mmm* using experimental data of [101].

The increase of the A atom size results in the increase of lattice parameters *a* (from 8.514 Å (LiCuO) up to 9.541 Å (RbCuO)) and *c* (from 3.809 Å (LiCuO) up to 5.812 Å (RbCuO)), whereas the latter one (parameter *c*) increases much more steeply due to chains of A tetrahedra stretched along the axis *c*. Accordingly, the Cu-Cu distances between $[Cu_4O_4]^{4-}$ rings in a stack increase significantly (by 2.00 Å: from 3.809 Å (LiCuO) up to 5.812 Å (RbCuO)), whereas the distance between stacks increases by as little as 1.30 Å (from 2.942 Å (LiCuO) up to 4.244 Å (RbCuO)). A different situation is observed for $Cu_4O_4$ rings ($d$(Cu-Cu) = 2.672 Å in LiCuO and 2.584 Å in RbCuO) – they shrink by 0.088 Å under effect of extending chains of A atoms: here the strength of metal bond in the ring increases, since the Cu-Cu distances in compounds of Na, K, and Rb become less than the sum of copper covalent radii (2.64 Å). The Cu-O bond length slightly decreases as well (from 1.85 (LiCuO) down to 1.82 (LiCuO) Å), which, according to BVS calculations, indicates to the increase of copper atoms charge from 1.26 up to 1.37.

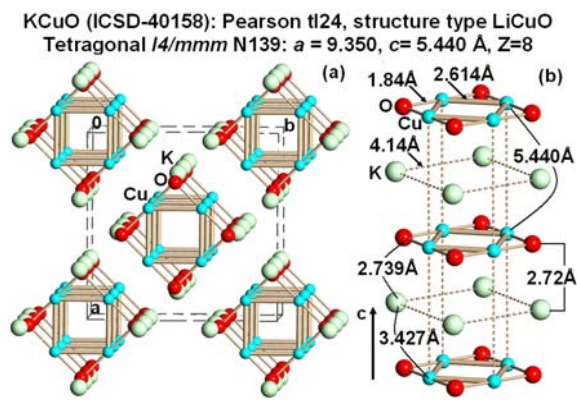

**Figure 12.** Crystal structure of KCuO [102] (a) and quasi-one-dimensional fragment – tube from square rings $[Cu_4O_4]^{4-}$, with potassium ions outside it.

To sum up, one can regulate, through complete or partial substitution of A atoms, the structural parameters of these compounds.

We assume that doping the compounds ACuO with holes through creation of vacancies in A atoms chains or substitution of a part of $O^{2-}$ ions by $N^{3-}$ ions could result in the emergence of superconductivity. The most suitable candidate to become a superconductor is the compound KCuO (figure 12(a)), since the Cu-Cu distances (5.440 Å) between $[Cu_4O_4]^{4}$ rings in KCuO stacks are close to those (5.443 Å) between copper atoms along the diagonals of squares in the $CuO_2$-plane of the superconductor $(Hg_{0.6}Tl_{0.4})Ba_2Ca_2Cu_3O_{8.33}$ [103]. Possibly, the compound LiCuO will have an ion conductivity, if one creates some deficit of $Li^+$ ions to induce the $Li^+$ ionic diffusion.

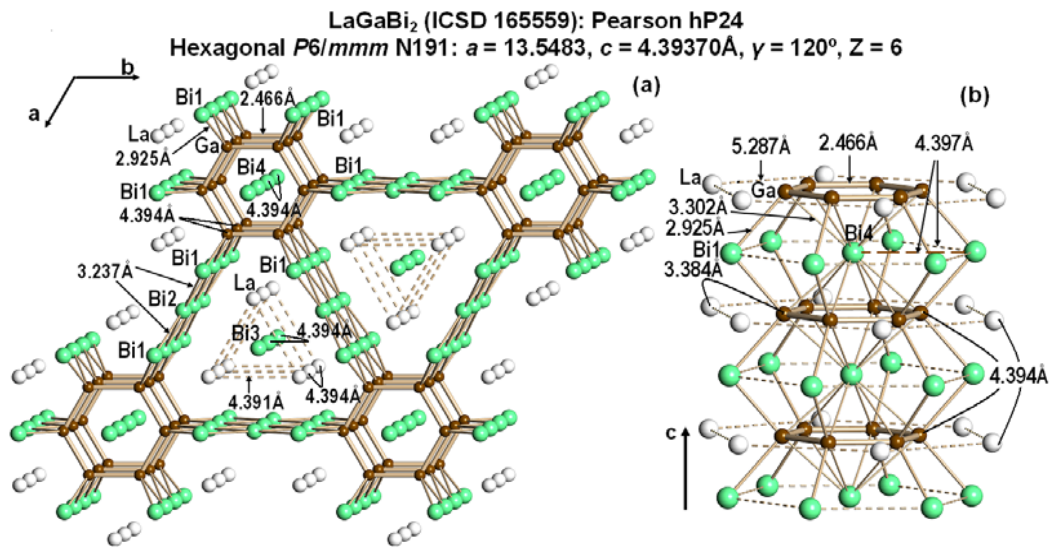

**Figure 13.** Crystal structure of LaGaBi$_2$ [104] (a) and quasi-one-dimensional fragment – tube from hexagonal Ga$_6$ rings, with Bi4$^{3-}$ ions inside it and La$^{3+}$ ions outside it.

*3.2.4. LaGaBi$_2$.* In the crystal structure of LaGaBi$_2$ (hexagonal, space group $P6/mmm$, $a$ = 13.548 Å, $c$ = 4.394(1) Å, $Z$ = 6) [104] (figure 13(a)), one can mark out three types of quasi-one-dimensional fragments along the $c$-direction: tubes from hexagonal Ga$_6$ rings, inside which Bi4 atoms are pressed between rings, columns of La$_6$ trigonal prisms centered by Bi3 atoms, and three-atom-wide Bi ribbons.

The Ga$_6$ rings are rigorously planar and have ideal bond angles of 120°. The Ga-Ga distances of 2.466 Å are virtually equal to the sum of covalent radii of Ga (2.44 Å). The Ga$_6$ rings are arranged in a one-dimensional stack along the $c$-direction (figure 13(b)), but the inter-ring distance is very long (equal to the $c$-parameter, 4.394 Å). The length of bonds between Ga and Bi4 (the atoms sandwiched between Ga$_6$ rings) of 3.302 Å exceeds significantly the sum of covalent radii (2.70 Å) of these atoms. In La$_6$ trigonal prisms all La-La distances (~4.39 Å) are larger (by ~0.84 Å), whereas the length of bonds La-Bi3 (3.355 Å) is, conversely, smaller (by 0.195 Å) than the sum of covalent radii. The Bi1-Bi2 distances in Bi ribbons (3.237 Å) are larger than the sum of covalent radii by 0.277 Å.

LaGaBi$_2$ is metallic with the resistivity decreasing only slightly with decreasing temperature [102]. In contrast to the closely related structure of La$_{13}$Ga$_8$Sb$_{21}$ (the onset of superconductivity below $T_c$ = 2.4 K) [105-107], it is not superconducting. However, there exists a principal difference between these structures in one fragment that, in our opinion, is responsible for the emergence of superconductivity. In La$_{13}$Ga$_8$Sb$_{21}$ similarly charged La$^{3+}$ ions are located inside and outside of hexagonal Ga$_6$ rings, whereas in LaGaBi$_2$ negatively charged Bi4 ions are located inside the tubes, and positively charged La ions are located outside them. Our studies of high-$T_c$ cuprates (see section 2) demonstrated that the charge of dielectric layers, between which the plane with carriers is located, was of primary importance for the emergence of superconductivity. In LaGaBi$_2$ we observed another case when dielectric fragments inside and outside the tube had opposite charges, which must hamper the emergence of superconductivity. To realize the latter situation, possibly, with high $T_c$ in compounds of the structural type LaGaBi$_2$ (whose formula within the frames of the elementary unit is La$_6$Ga$_6$Bi$_{12}$), we suggest to carry out a number of substitution and dope them with holes: (i) substitute Bi4 atoms located inside the tube by atoms of La or other rare earth element (Re); (ii) substitute Ga atoms by those of d-element Fe that must be possible, since Ga and Fe form isomorphic mixtures [108, 109]; (iii) substitute a part of La ions by calcium ions. As a result, the compound formula will become (Re$_{7-x}$Ca$_x$)Fe$_6$Bi$_{11}$ in the LaGaBi$_2$-type. One can also attempt to obtain the compound (Re$_{7-x}$Ca$_x$)Fe$_6$Sb$_{11}$ in this structural type.

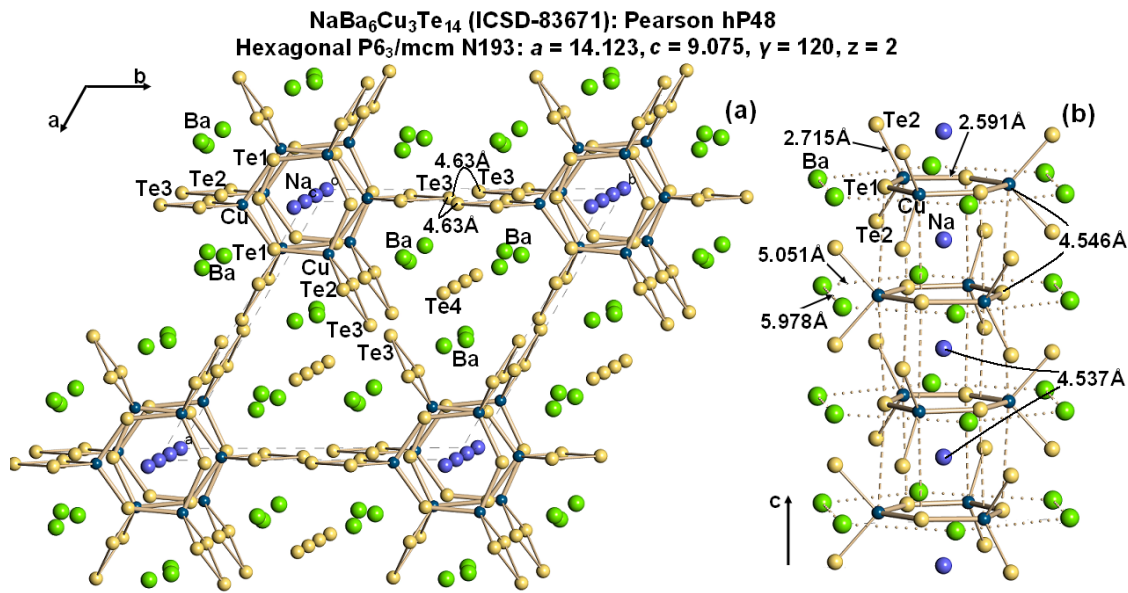

**Figure 14.** Crystal structure of $NaBa_6Cu_3Te_{14}$ [110] (a) and quasi-one-dimensional fragment – tube from six-membered $Cu_3Te_3$ rings with $Na^+$ ions inside it and $Ba^{2+}$ ions outside it.

*3.2.5. $NaBa_6Cu_3Te_{14}$.* The crystal structure of $NaBa_6Cu_3Te_{14}$ (hexagonal, space group $P6_3/mcm$ N193, $a$ = 14.123(2) Å, $c$ = 9.075(1) Å, $Z$ = 2) [110] (figure 14(a)) contains a quasi-one-dimensional fragment which satisfies our requirements in structural terms. Its structure is based on the $[Cu_3Te_3(Te_3)_3]^{9-}$ clusters. These clusters comprise six-membered $Cu_3Te_3$ rings located on a $\bar{6}2m$ site, they must be strictly planar. The Cu-Te1 bonds, Te1-Cu-Te1 and Cu-Te1-Cu angles in the $Cu_3Te_3$ ring are 2.59 Å, 109.4° and 130.6°, respectively. Each copper atom is coordinated in a bidentate fashion to the $Te_3^{2-}$ ligand to form the four-membered $CuTe_3$ ring, which is also planar and lies perpendicularly to the $Cu_3Te_3$ ring. The $[Cu_3Te_3(Te_3)_3]^{9-}$ clusters stack along the $c$ axis in a staggered arrangement and encapsulate $Na^+$ ions inside the column stack. The distances between $Cu_3Te_3$-rings in a stack are equal to 4.537 Å ($c$/2). The distances Na-Cu and Na-Te between Na atoms and $Cu_3Te_3$ rings are equal to 3.540 Å and 3.333 Å, respectively. The clusters linked this way and shaped like a paddle-wheel form a highly negatively charged, one-dimensional column of $[NaCu_3Te_3(Te_3)_3]^{8-}$ along the $c$ axis. These columns are surrounded by six columns of divalent barium cations. In other words, this structure contains tubes from six-membered $Cu_3Te_3$ rings, $Na^+$ ions are located inside the tubes, $Ba^{2+}$ ions are located outside them (see figure 14(b)). Elsewhere in the lattice discrete $Te^{2-}$ ions (Te4) are each surrounded by six $Ba^{2+}$ ions in a trigonal prismatic arrangement Therefore, the structure of this telluride is best described by the formula $Ba_6[NaCu_3Te_3(Te_3)_3]Te_2$, as was suggested in [111].

According to the data of measurements [110], this telluride is a p-type semiconductor that is explained in [110, 111] by the existence of ionic Te⋯Te contacts, which are shortened as compared to the sum of the van der Waals radii.

One can assume that doping six-membered $Cu_3Te_3$-rings with holes would result in the emergence of superconductivity in $NaBa_6Cu_3Te_{14}$. Doping can be realized through substitution of $Ba^{2+}$ ions by $K^+$ ions: here one can leave $Na^+$ ions in the same position and obtain the compound $NaBa_{6-x}K_xCu_3Te_{14}$ or substitute them completely by $K^+$ ions and obtain the compound $K_{1+x}Ba_{6-x}Cu_3Te_{14}$. There is also one more way – substitution of $Te^{2-}$(preferably Te4 (site 4d)) by $Sb^{3-}$.

## 4. Conclusions

We believe that superconductors with the quasi-one-dimensional crystal structure based on the Ginzburg sandwich, comprising a layered structure dielectric–metal–dielectric, convoluted into a tube can have high temperatures of the superconducting transition. The dimensionality reduction from three to two dimensions enables one to increase $T_c$ from 30 K (in $Ba_{1-x}K_xBiO_3$) up to 138 K (in $HgBa_2Ca_2Cu_3O_{8+\delta}$, doped with Tl). These tubes of the sandwich type comprise conducting tubes or those from conducting rings with dielectric inside and outside them. The system metal–dielectric can be also suitable only when dielectric is present only inside or only outside of a conducting tube or that built of conducting rings. This assumption is based on the fact that crystal chemistry factors creating the conditions for the emergence of superconductivity and determining the value of $T_c$ in $A$-$CuO_2$-$A$ sandwiches of layered superconducting cuprates and tubes of the sandwich type must be essentially similar. Besides, we managed to mark out similar quasi-one-dimensional fragments in known low-temperature superconductors that, in our opinion, are responsible for the emergence of superconductivity, but do not to full degree comply with criteria determining high transition temperatures.

In the inorganic crystal structure database (ICSD) we identified five compounds ($La_4BaCu_4FeO_{12.96}$, $Ba_9Cu_7O_{15}Cl_2$, KCuO, $LaGaBi_2$, and $NaBa_6Cu_3Te_{14}$) whose structures mainly correspond to the determined crystal chemistry conditions, at which the emergence of superconductivity with high superconducting transition temperatures is possible. These compounds could be the parents of new families of inorganic superconductors with quasi-one-dimensional structure upon doping, substitutions, or modifications in variants we have suggested. Nevertheless, the history of superconductivity studies demonstrates that although Ginzburg once managed to predict the structural type of high-temperature superconductors, their chemical composition is still rather ambiguous (researchers mostly reveal novel HTSC at random). That is why the compounds containing tubes of the sandwich-type must undergo various substitution, with subsequent tests related to superconductivity.